\documentclass[%
 aip,
 amsmath,amssymb,
 reprint,%
]{revtex4-1}

\usepackage{graphicx}
\usepackage{amsmath}
\usepackage{amsfonts}
\usepackage{dcolumn}
\usepackage{bm}

\usepackage[utf8]{inputenc}
\usepackage[T1]{fontenc}
\usepackage{mathptmx}
\usepackage{etoolbox}
\usepackage{xcolor}
\usepackage{braket}
\usepackage{xfrac}
\usepackage{hyperref}

\makeatletter
\def\@email#1#2{%
 \endgroup
 \patchcmd{\titleblock@produce}
  {\frontmatter@RRAPformat}
  {\frontmatter@RRAPformat{\produce@RRAP{*#1\href{mailto:#2}{#2}}}\frontmatter@RRAPformat}
  {}{}
}%
\makeatother
\begin{document}

\preprint{AIP/123-QED}

\title{Topological Hall effect in nonlinear optics}
\author{Soumik Nandi}
\affiliation{School of Physical Sciences, National Institute of Science Education and Research, Odisha 752050, India}
\affiliation{Homi Bhabha National Institute, Training School Complex, Anushaktinagar, Mumbai 400094, India}

\author{Arannya Ghosh}%
\affiliation{Optics and Photonics Centre, Indian Institute of Technology Delhi, Delhi 110016, India}

\author{Ashok K Mohapatra}
\affiliation{School of Physical Sciences, National Institute of Science Education and Research, Odisha 752050, India}
\affiliation{Homi Bhabha National Institute, Training School Complex, Anushaktinagar, Mumbai 400094, India}

\author{Ritwick Das}
\email{dasritwick@opc.iitd.ac.in}
\affiliation{Optics and Photonics Centre, Indian Institute of Technology Delhi, Delhi 110016, India}

\date{\today}

\begin{abstract}
We present an experimental evidence of \emph{topological} Hall-effect in an all-optical third-order nonlinear optical process via spatial symmetry-breaking in pseudo-spin textures created by a spatially-structured pump laser beam. The experimental configuration consists of a moderately-focused pump laser beam undergoing a parametric interaction with an organic solvent (toluene) and an off-resonant laser beam probes the non-trivial spatial magnetization textures created by the pump beam. The phase-profile of the transmitted probe beam is extracted using phase-retrieval algorithms for ascertaining the topological charge which is shown to be consistent with the estimation of Berry's curvature that we obtain via paraxial approximation-based modeling of third-order nonlinear interaction. 
\end{abstract}

\maketitle

Particle spin dynamics provides one of the richest frameworks for exploring quantum geometry and its manifestations \cite{OECKL2001233}. Fundamentally, the relativistic formalism of quantum mechanics by Dirac necessitated the existence of spin and its dynamical interaction in presence of magnetic field \cite{10.1063/1.3062610}. The unwitting first experimental evidence of spin was revealed through the famous Stern-Gerlach experiment \cite{gerlach_experimentelle_1922} which is followed by innumerable tests on this degree of freedom \cite{Karpa2006,RevModPhys.85.299}. In condensed-matter physics, spin-dynamics and its geometrical interpretation forms the foundational principles of \emph{spintronic} devices \cite{RevModPhys.76.323}. 
Such configurations are indispensable ingredients of non-volatile magnetic random access memory (MRAM) \cite{10.1063/1.369932}, magnetometers \cite{leitao_enhanced_2024}, automotive braking systems \cite{8951547}, and racetrack memory \cite{doi:10.1126/science.1145799}, with significant applications in perimeter defense systems \cite{7559746}. By leveraging the features of scattering-immune spin transport via the creation of topologically non-trivial spin-polarized states, 
the next generation of spintronic devices is expected to exhibit a significant reduction in power consumption and improved device functionalities \cite{10.1063/5.0218280}.
All-optical analogue of topological Hall-effect (THE) and the existence of optical Skyrmions facilitate greater control of the \emph{pseudo}-spin transport mechanism via tuning suitable optical degree of freedom \cite{10.1063/5.0241546}. Conventionally, the geometric character in polarization optics is introduced via adopting suitable gauge symmetries in the paraxial dynamics in optically birefringent systems \cite{10.1063/1.4938003}. By virtue of the fact that nonlinear optical configurations exhibit perturbative dynamics naturally, they provide a fertile ground for exploring geometrical manifestations and topological connections in photonics. In a seminal work by Karnieli \emph{et al.}, an equivalence was drawn between a $\chi^{(2)}$-based frequency conversion process (in \emph{frequency-state} basis) and the dynamics of a spin-$\sfrac{1}{2}$ particles in a magnetizing field \cite{Karnieli2021}. Such an equivalence paves the route for developing analogous all-optical spintronic functionalities such as the \emph{"Giant Magneto Resistance"} (GMR) effect as well as the creation of topologically non-trivial 3D magnetization textures \cite{Izhak:24}. It is worthwhile to mention that an all-optical setting for topology-assisted \emph{pseudo}-spin transport exhibits an unique feature in terms of extremely broad operational  bandwidth. An integrated photonics based approach further facilitates preserving the state fidelity (low coupling losses) and hence, high signal-to-noise ratio \cite{10.1063/5.0218564}. Amongst the broad category of nonlinear optical processes, $\chi^{(3)}$-led self-action and four wave-mixing (FWM) effects, are the most prevalent. In particular, the dynamics of self-action effects, such as self-focusing and filamentation, tend to exhibit intricate topologically non-trivial manifestations which are revealed through complex spatial beam structures \cite{PhysRevA.94.023805}. In this letter, we present an equivalent THE in a $\chi^{(3)}$-mediated nonlinear optical interaction. The pump-probe based experimental measurements reveal the existence of \emph{pseudo}-spin basis modes which exhibit a symmetric transverse separation in presence of a `experimentally-controllable' all-optical gauge-field. Through an analytical formalism, we show that a spatial asymmetry in the pump laser beam leads to topologically non-trivial, gauge-dictated, synthetic magnetization textures which is responsible for probe beam splitting thereby, mimicking the THE. Interestingly, the dynamical interaction adheres to a $\mathcal{PT}$-symmetric non-Hermitian system that exhibits a real-valued eigenvalue spectrum \cite{PhysRevLett.80.5243,PhysRevLett.89.270401} for $\chi^{(3)}~< 0$. Overall, the results derive their novelty via the demonstration for proof-of-principle extension of THE-like dynamics to dynamically non-Hermitian counterparts.

Let us consider the propagation of a strong light beam at frequency $\omega_{p}$ along $z$ direction in a $\chi^{(3)}$ nonlinear medium. In a scalar approximation, we define $\ket{\psi_{p}}=\frac{1}{2}\left[a_{p}(\vec{r})e^{i(\omega_{p}t-k_{p}z)}+c.c\right]$ where $k_{p}$ is the wavevector of the beam. The strong pump beam spontaneously generates two co-propagating fields \emph{viz.} $\ket{\psi_{+}}$ and $\ket{\psi_{-}}$ which are governed by the linearized nonlinear Schr\"{o}dinger's equation (NLSE) in the paraxial framework and is given by \cite{boyd_nonlinear_2008},
\begin{equation}
    \nabla_{T}^2 a_{+}+2ik_{p}\frac{\partial a_{+}}{\partial z}=-\frac{6\chi^{(3)}\omega_{p}^2}{c^2}\vert a_{p}\vert^2a_{+}-\frac{3\chi^{(3)}\omega_{p}^2}{c^2}a_{p}^2e^{2i\zeta}a_{-}^{*}
\end{equation}

\begin{equation}
    \nabla_{T}^2 a_{-}+2ik_{p}\frac{\partial a_{-}}{\partial z}=-\frac{6\chi^{(3)}\omega_{p}^2}{c^2}\vert a_{p}\vert^2a_{-}-\frac{3\chi^{(3)}\omega_{p}^2}{c^2}a_{p}^2e^{2i\zeta}a_{+}^{*}
\end{equation}
where we have defined $\ket{\psi_{\pm}}=\frac{1}{2}\left[a_{\pm}(\vec{r})e^{i(\omega_{p}t-k_{p}z)}+c.c\right]$, $\zeta=\frac{\omega_{p}}{c}\int_{0}^{z} [n_{p}+n_{2}I_{p}(\vec{\rho},z')]\,dz'$ and $n_{2}$ and $I_{p}$ represents nonlinear refractive index of medium and on-axis peak intensity of pump beam respectively. Using an unitary transformation that is carried out by the operator $\mathbb{P} = diag~(e^{i\zeta}, e^{-i\zeta})$, $z$-independent dynamical Eqs. are expressed as \cite{Nandi:25, Karnieli2021}
\begin{equation}
    i\frac{\partial}{\partial z}\ket{\Phi}=\left[\frac{(\vec{p}_{T}-\vec{\mathcal{A}})^2}{2\bar{m}}-\vec{\sigma}\cdot\vec{\mathcal{M}}\right]\ket{\Phi} = \hat{H}\ket{\Phi} 
    \label{eq:3}
\end{equation}
where $\ket{\Phi}=\mathbb{P}^\dagger(a_{+}~~a_{-}^{*})^{T}$ defines the spinor in rotated frame. $\vec{p}_{T}=-i\vec{\nabla}_{T}=-i(\hat{x}\frac{\partial}{\partial x}+\hat{y}\frac{\partial}{\partial y})$ is the equivalent momentum operator, $\bar{m} = k_p\sigma_z$ is equivalent mass operator and $\vec{\sigma}\equiv[\sigma_{x},\sigma_{y},\sigma_{z}]$ is the triad of Pauli spin matrices. $\vec{\mathcal{A}}=-\vec{\nabla}_{T}\zeta\sigma_{z}$ is gauge-dictated magnetic vector potential, $\vec{\mathcal{M}}=\mathcal{M}_{0}\hat{\mathcal{M}}=\Omega\sin(2\varphi)e^{-i(2m+1)\pi/2}\hat{x}+\Omega\cos(2\varphi)e^{i(2m+1)\pi/2}\hat{y}+ (k_p+\Omega)\hat{z}$ is the synthetic magnetization created by the pump beam where $\Omega=\frac{\omega_{p}}{c}n_{2}I_{p}=\frac{3\chi^{(3)}\omega_{p}^2}{2k_{p}c^2}|\mathbb{A}_{0}|^2$ and $a_{p}=\mathbb{A}_{0}e^{i\varphi(x,y)}$ (where $\varphi$ is real). It is worth pointing out that the dynamical evolution of $\ket{\psi_{\pm}}$ states as depicted by $\hat{H}$ has a non-Hermitian character with a preserved $\mathcal{PT}$-symmetry when $\chi^{(3)} < 0$ \cite{Nandi:25}. From an alternative perspective, the approach provides a platform for realizing complex all-optical synthetic magnetization textures $\left(\vec{\mathcal{M}}(x,y)\right)$ via altering the structural symmetry of the pump beam. The propagation of \emph{pseudo}-spin states ($\ket{\psi_{\pm}}$) through the created magnetization textures could be precisely controlled in space that is reflected through restructuring of probe beam. Eq. (\ref{eq:3}), therefore, provides a broad framework for exploring a quantum two-level dynamics in a purely classical $\chi^{(3)}$-led self-action  effect. In order to reveal the impact of this approach, we use a transformation $\ket{\Phi}=D\ket{\Phi^{'}}$ which essentially aligns the local magnetization direction ($\hat{\mathcal{M}}$) along $\hat{z}$ where $D$ is defined as
\begin{equation}
    D=\cos{\frac{\theta}{2}}\mathbb{I}-i\sin{\frac{\theta}{2}}(\vec{\sigma}\cdot\hat{\phi})
    \label{eq:4}
\end{equation}
where $\theta=\arccos(\hat{\mathcal{M}}\cdot\hat{z})$ is polar angle and $\phi=tan^{-1}(\mathcal{M}_{y}/\mathcal{M}_{x})$ is azimuthal angle with $\mathbb{I}$ defining the identity operation. Consequent upon this, Eq. (\ref{eq:3}) transforms to
\begin{equation}
    i\frac{\partial}{\partial z}\ket{\Phi^{'}}=\left[\frac{(\vec{p}_{T}-\vec{\mathcal{A}^{'}})^2}{2\bar{m}}+V\right]\ket{\Phi^{'}}
    \label{eq:5}
\end{equation}
where, the transformed gauge-dictated vector and scalar potentials are $\vec{\mathcal{A}^{'}}=iD^{\dagger}\vec{\nabla}_{T}D+D^{\dagger}\vec{\mathcal{A}}D$ and $V=-iD^{\dagger}\frac{\partial}{\partial z}D-\mathcal{M}_{0}\sigma_{z}$ respectively (see SI document). In this rotated frame, we only consider terms proportional to $\sigma_{z}$ and in the adiabatic limit, the transformed potentials are $\vec{\mathcal{A}^{'}}=\{\cos{\theta}~\vec{\nabla}_{T}(\phi/2-\zeta)\}\sigma_{z}-\frac{\vec{\nabla}_{T}\phi}{2}\sigma_{z}$ and $V=-\mathcal{M}_{0}\sigma_{z}$ \cite{10.1063/1.4870695,Karnieli2021}. Hence, we define the synthetic magnetic field ($\vec{\mathcal{B}}$) and synthetic electric field as ($\vec{\mathcal{E}}$) as below
\begin{equation}
\begin{split}
\vec{\mathcal{B}}&=\vec{\nabla}_{T}\times\vec{\mathcal{A}^{'}}=-\left(\frac{k_{p}\Omega}{\mathcal{M}_{0}^3}\vec{\nabla}_{T}\Omega\times\vec{\nabla}_{T}\varphi\right)\sigma_{z} \\
    & = -\frac{1}{2}\hat{z} \left[\hat{\mathcal{M}}\cdot\left(\frac{\partial\hat{\mathcal{M}}}{\partial x}\times\frac{\partial\hat{\mathcal{M}}}{\partial y}\right)\right]
    \end{split}
    \label{eq:6}
\end{equation}
\begin{equation}
\begin{split}
    \vec{\mathcal{E}}&=-\vec{\nabla}_{T}V-\frac{\partial\vec{\mathcal{A}^{'}}}{\partial z}\\
    &=\left[\vec{\nabla}_{T}\mathcal{M}_{0}+\left(\frac{\mathcal{M}_{z}}{\mathcal{M}_{0}}\right)\vec{\nabla}_{T}\Omega+\frac{k_{p}\Omega}{\mathcal{M}_{0}^3}\frac{\partial\Omega}{\partial z}\vec{\nabla}_{T}(\varphi+\zeta)\right]\sigma_{z}
\end{split}
\label{eq:7}
\end{equation}
Since, a geometric interpretation of $\vec{\mathcal{E}}$ is hinged on $z$-dependence of $\vec{\mathcal{A}}'$ ($\propto~\Omega$), we focus on $\vec{\mathcal{B}}$ which could lead to geometric phases being acquired by the optical state $\ket{\Phi'}$. It is apparent from Eq. (\ref{eq:6}) that a Gaussian pump beam $\left(a_{p}^{(G)}\sim e^{-\frac{r^2}{w_p^2}}\right)$ manifests into $\vec{\mathcal{B}}=0$. On the other hand, a spatial asymmetry in the pump beam structure gives rise to the possibility of topological bands. We adopt a simple route for breaking radial symmetry in the pump beam by considering an elliptic Gaussian (EG) pump beam given by $a_{p}^{(EG)}=\mathbb{A}_{0}~exp{\left(-\frac{x^2}{w_{x}^2}-\frac{y^2}{w_{y}^2}\right)}~exp{\left(-ik_{p}\frac{x^2}{R_{x}}-ik_{p}\frac{y^2}{R_{y}}\right)}$ where [$w_{x}$, $w_{y}$] and [$R_{x}$, $R_{y}$] are beam waists and radius of curvatures along $\hat{x}$ and $\hat{y}$ directions respectively. $\vec{\mathcal{B}}$, in such a situation, takes a form,
\begin{equation}
\begin{split}
\vec{\mathcal{B}}&\approx(-\hat{z})\frac{8\Omega^2}{k_p}xy\left(\frac{1}{w_{y}^2R_{x}}-\frac{1}{w_{x}^2R_{y}}\right)\sigma_{z}
\end{split}
\label{eq:8}
\end{equation}
where $\Omega$ is redefined as $\Omega = \Omega_0 e^{-\left(\frac{2x^2}{w_{x}^2}+\frac{2y^2}{w_{y}^2}\right)}$ with $\Omega_0 =\frac{3\chi^{(3)}\omega_{p}^2}{2k_{p}c^2}|\widetilde{\mathbb{A}}_{0}|^2$. In writing Eq. \eqref{eq:8}, we have assumed $k_p >> |\Omega_{0}|$ which represents a massive fermion dynamics. By virtue of non-zero $\vec{\mathcal{B}}$, the Lorentz force ($\vec{\mathcal{F}} = q_e \vec{v}_T\times\vec{\mathcal{B}}$) on probe photons will be,
\begin{equation}
\begin{split}
    \vec{\mathcal{F}} = \kappa_0e^{\left(-\frac{4x^2}{w_{x}^2}-\frac{4y^2}{w_{y}^2}\right)}\left[x\left(1-\frac{8y^2}{w_{y}^2}\right)\hat{x}-y\left(1-\frac{8x^2}{w_{x}^2}\right)\hat{y}\right]
\end{split}
\label{eq:9}
\end{equation}
where $\kappa_0 =\frac{8iq_{e}\Omega_{0}^2}{k_{p}^2}\left(\frac{1}{w_y^2 R_x}-\frac{1}{w_x^2 R_y}\right)$. Here $q_e = \pm 1$ is the emergent charge corresponding to each pseudo-spin basis constituting the probe beam and bears a sign determined by pseudo-spin orientation with respect to the local magnetization ($\vec{\mathcal{M}}$) axis. We consider elliptic Gaussian beam with ellipticity $\frac{w_x}{w_y} = 0.90,~0.40,~0.12$ as shown in  Fig. 1(a)-(c). Corresponding to this, we present the force field distribution $\left(\frac{\vec{\mathcal{F}}}{\kappa_0}\right)$ in Fig. 1(d)-(f). It is apparent that $|\vec{\mathcal{F}}|$ exerts an azimuthally symmetric force on probe photons when pump has small ellipticity ($0.9$). An increase in pump beam ellipticity results in $|\vec{\mathcal{F}}|$ localized close to $y$-axis and directed in $\pm x$-direction, thereby depleting the probe photon density on the $y$-axis. Therefore, it is expected that the probe beam would localize in the $x-y$ plane away from the $y$-axis as a consequence of increase in pump beam ellipticity. Additionally, it is worth noting that the magnitude of $\vec{\mathcal{F}}$ is stronger at higher pump powers and therefore, the impact of pump beam ellipticity would be more discernible. In order to gain a deeper insight, we solved Eq. (\ref{eq:5}) numerically via beam propagation method and used experimental values of physical quantities which is described in the next section. By considering two examples where $w_{x} = 500~\mu m,~w_y = 1200~\mu m$ and $w_{x}= 150~\mu m,~w_y = 1200~\mu m$ for the nonlinear medium of length $T$, we simulated the probe beam profile at the exit face ($z = +\frac{T}{2}$) at different values of incident pump powers ($P_p$) and shown in Fig. 2(a)-(d) and Fig. 2(e)-(h) respectively. A comparison between Fig. 2(a)-(d) and Fig. 2(e)-(h) shows that the probe beam distortion is elucidated more when pump beam ellipticity is high. Also, the probe beam distorts more at higher pump powers for a given pump ellipticity. The conclusions drawn from the simulated $\vec{\mathcal{F}}$ in Fig. 1(d)-(f) predicts a similar variation in the transmitted probe beam. In fact, there is a distinct appearance of two lobes in the simulated probe beam for the pump beam ellipticity of $0.12$ (for $P_p = 500~mW$) and the centroids of the two lobes are separated by $\approx 0.16~mm$ at $T =\pm\frac{ T}{2}$.   

\begin{figure}
    \centering
    \includegraphics[width=\linewidth]{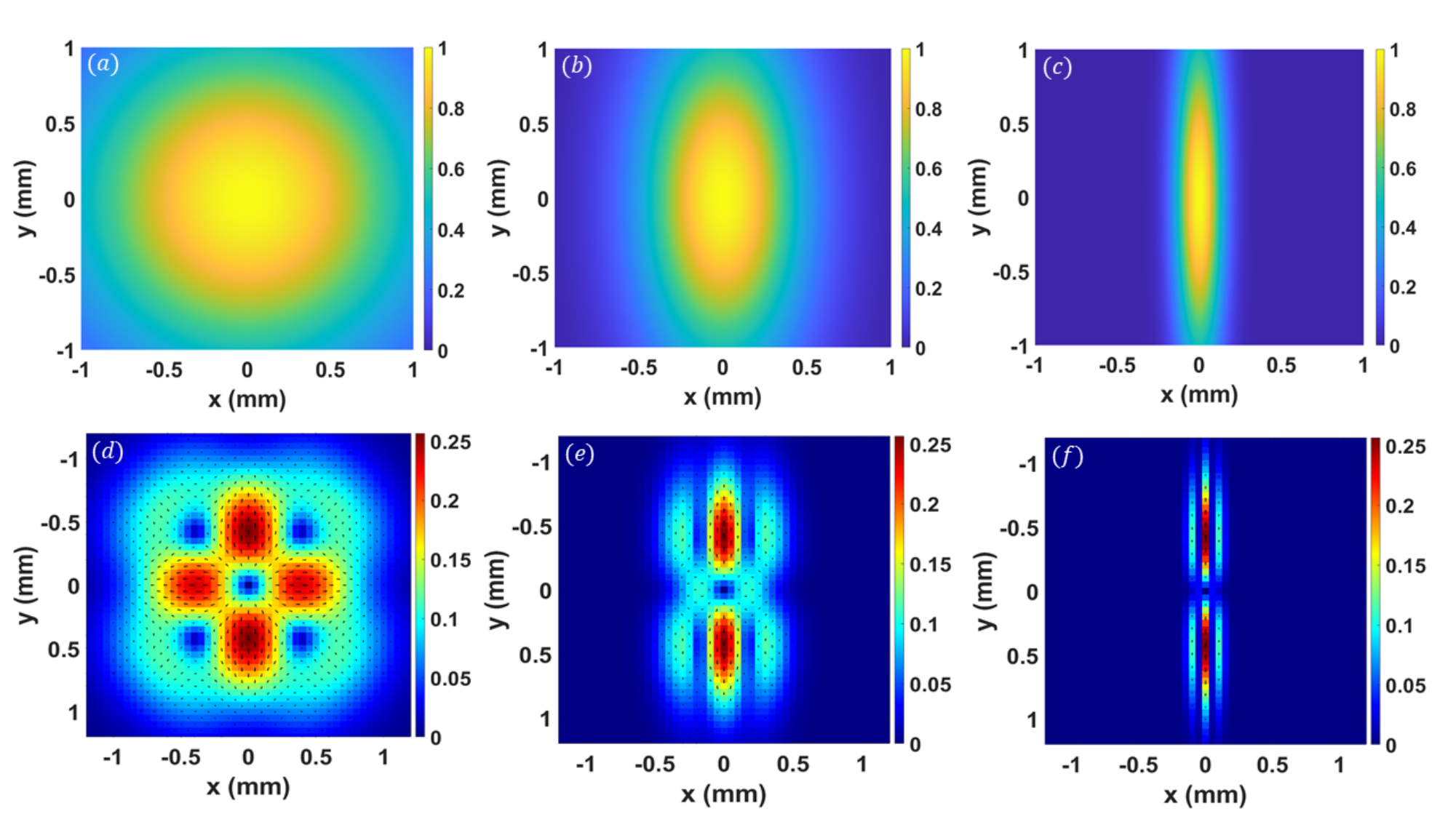}
    \caption{(a)-(c) represents elliptic pump beam profiles for $w_{x} = 1100~\mu m, ~500~\mu m, 150~\mu m$ and $w_{y} = 1200~\mu m$ corresponding to ellipticity of $0.90$ (a), $0.40$ (b), $0.12$ (c) respectively. (d)-(f) represents a force  field ($|\vec{\mathcal{F}}|$) intensity map for pump beam with ellipticity of $0.90$ (a), $0.40$ (b), $0.12$ (c) respectively.}
    \label{fig1}
\end{figure}
\begin{figure}
    \centering
    \includegraphics[width=\linewidth]{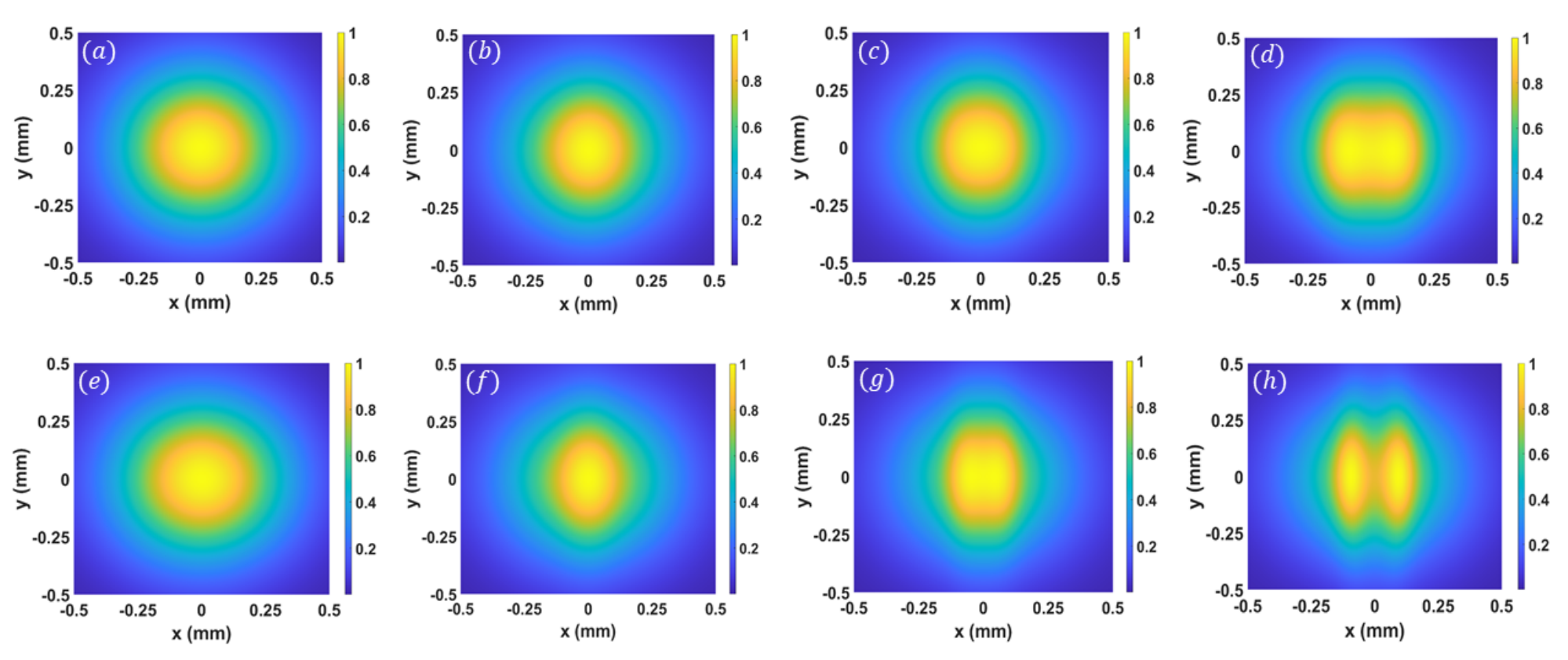}
    \caption{(a)-(d) represents simulated probe beam profiles at the exit face nonlinear medium ($z = +\frac{T}{2}$) for $w_{x} = ~500~\mu m$ and $w_{y} = 1200~\mu m$ at average pump beam power of (a) $30~mW$, (b) $150~mW$, (c) $300~mW$, (d) $500~mW$ respectively. (e)-(h) represents simulated probe beam profiles at the exit face nonlinear medium ($z = +\frac{T}{2}$) for $w_{x} = ~150~\mu m$ and $w_{y} = 1200~\mu m$ at average pump beam power of (a) $30~mW$, (b) $150~mW$, (c) $300~mW$, (d) $500~mW$ respectively.}
    \label{fig2}
\end{figure}

The analytical framework depicts that the transmitted probe beam will acquire a geometric phase in presence of a non-vanishing $\vec{\mathcal{B}}$ field which is given by
\begin{equation}
\begin{split}
\Theta^{(EG)}&=\frac{1}{2\pi}\iint_{S}\vec{\mathcal{B}}\cdot\vec{dS}\\
    &=\frac{4}{\pi k_{p}}\left(\frac{1}{w_{x}^2R_{y}}-\frac{1}{w_{y}^2R_{x}}\right)\iint_{S}\Omega^2 xy\,dx\,dy
    \label{eq:10}
\end{split}
\end{equation}
In order to simplify Eq. (\ref{eq:10}), we choose polar coordinates \emph{i.e.}  $x=\rho\cos{\frac{\phi}{n}}$ and $y=\rho\sin{\frac{\phi}{n}}$ where $n$ defines the number of symmetry axis in $\vec{\mathcal{B}}(x,y)$. In case of an elliptic pump beam, $n = 2$ by virtue of an inherent two-fold spatial symmetry-axis. Therefore, for $\rho = 0~\rightarrow~\rho = \infty$, Eq. (\ref{eq:10}) simplifies to    
\begin{equation}
    \Theta^{(EG)}=\frac{\Omega_0^2}{32\pi k_p}\left(\frac{1}{w_{x}^2R_{y}}-\frac{1}{w_{y}^2R_{x}}\right) \int_{\phi_1}^{\phi_2} \frac{\sin\left(\phi\right)}{\alpha(\phi)^2} \, d\phi
    \label{eq:11}
\end{equation}
where $\alpha(\phi) = \frac{\cos^2\left(\frac{\phi}{2}\right)}{w_x^2} + \frac{\sin^2\left(\frac{\phi}{2}\right)}{w_y^2}$. This has a non-zero contribution to the integration in Eq. (\ref{eq:11}) when $\phi$ goes from $\phi_1 = 0 \rightarrow \phi_2 = 2\pi$. Therefore, $\Theta^{(EG)}_+ = \theta_0$ when $\phi_1 = 0~\rightarrow~\phi_2 = \pi$ and $\Theta^{(EG)}_- = -\theta_0$ when $\phi_1 = \pi~\rightarrow~\phi_2 = 2\pi$ respectively. In other words, the transmitted probe beam lobes (in Fig. 2(h)) have quantized geometric phase bearing opposite sign. Therefore, the two lobes constitute two topologically non-trivial spatial bands separated by a gap governed by intensity of the pump beam. The manifestation is reminiscent of topological Hall effect (THE) in an all-optical configuration where the (synthetic) magnetization texture is created and controlled by intensity and spatial structure of the pump beam.   


\begin{figure}
    \centering
    \includegraphics[width=\linewidth]{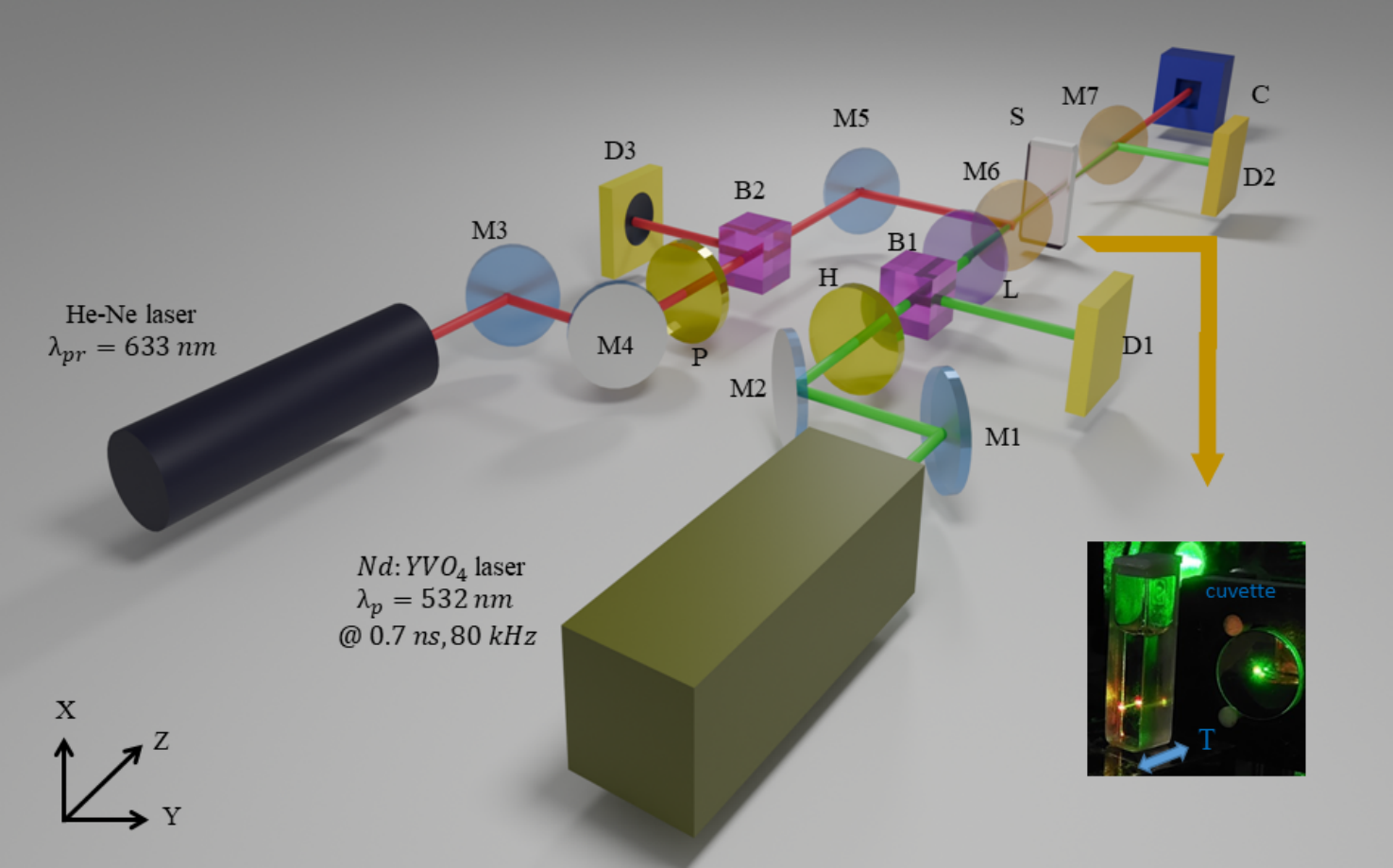}
    \caption{shows the collinear pump-probe based experimental configuration for observing topological Hall effect (THE) in an asymmetric synthetic magnetic field ($\vec{\mathcal{B}}$). B1: Polarizing beam-splitter; B2: Non-polarizing beam-splitter; S: sample; D1-D3: Power meter; H: Half-wave plate; P: Polarizer; M1-M5: Steering mirror; M6 \& M7: Dichroic mirrors for realizing collinear arrangement; L: cylindrical lens;C: CCD camera. \textit{Inset}: $T = 1~cm$ thick cuvette filled with toluene that exhibits $\chi^{(3)} < 0$ (defocusing) for $700~ps$ pulsed laser centered at $532~nm$ wavelength.}
    \label{fig3}
\end{figure}
The experimental configuration is a collinear pump-probe setup realized using a frequency-doubled $Nd: YVO_4$ pulsed laser delivering pulses of $\sim 1~ns$ width at $532~nm$ central wavelength acting as a pump beam. The pump pulses are linearly polarized $TEM_{00}$ Gaussian profile. A combination of half wave plate (HWP) and polarizing beam splitter (PBS) is used for controlling the incident pump power. The pump beam passes through a plano-convex cylindrical lens (CL) of appropriate focal length ($f$) which creates an elliptic beam profile at the focus. The beam ellipticity and dimensions are estimated using a knife-edge based scanning beam profiler (NanoScan, Ophir, USA). We used an organic solvent toluene as a nonlinear medium which was housed in a $T = 1~cm$ thick cuvette and positioned at the pump beam focus ($z = 0$) such that the laser beam entry and exit faces are given by $z=\pm \frac{T}{2}$. The probe beam is the red beam at $633~nm$ from a He-Ne laser and both the beams were made collinear using a dichroic mirror before they are incident on toluene. Another dichroic mirror was employed to separate the two beams. A CCD camera of the pixel size of $4.4~\mu m\times4.4~\mu m$ (GRAS 2.0, Point Grey, USA) was used for recording the probe beam structure and it was positioned at a distance $\approx30~cm$ from the cuvette. 

\begin{figure}
    \centering
    \includegraphics[width=\linewidth]{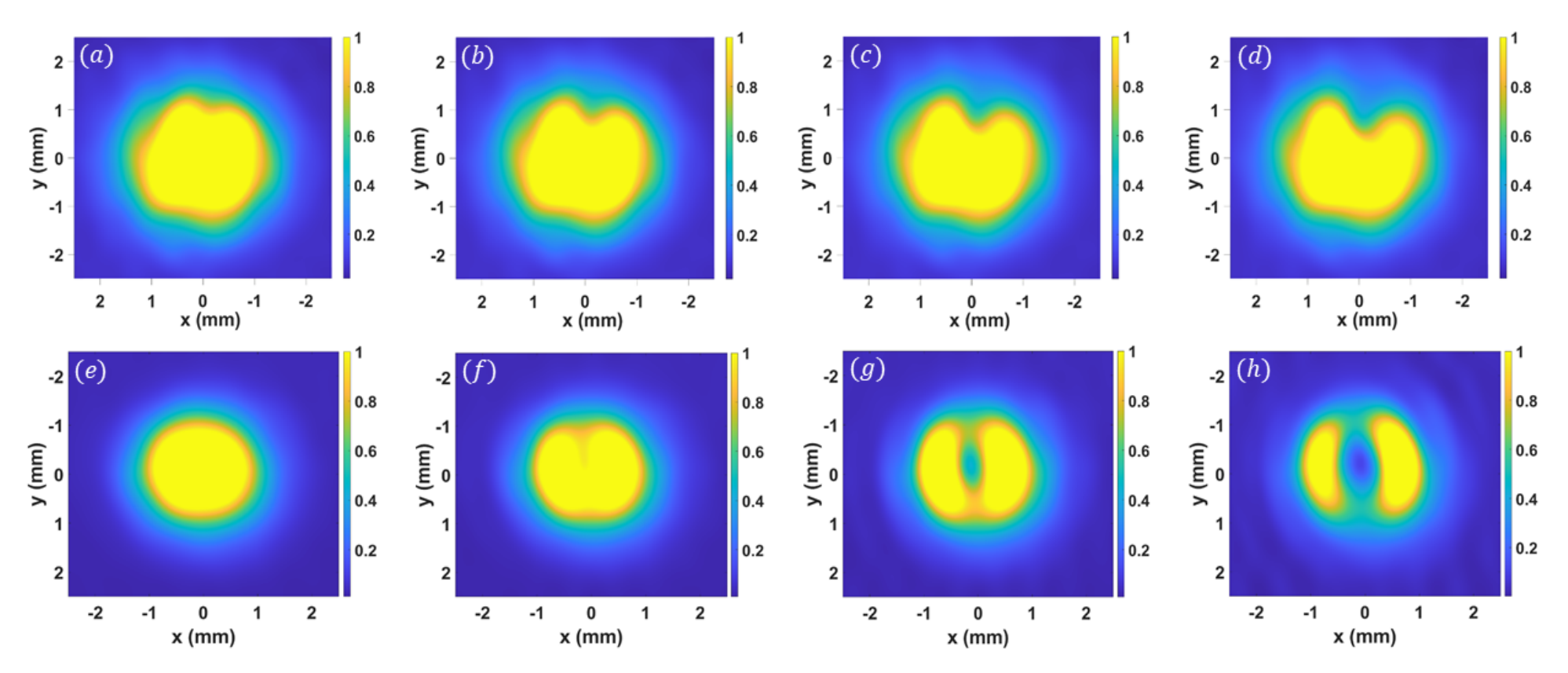}
    \caption{(a)-(d) represents experimentally measured probe beam profiles at the far field  ($z = \frac{T}{2}+30~cm$) for $w_{x} = ~500~\mu m$ and $w_{y} = 1200~\mu m$ at average pump beam power of (a) $30~mW$, (b) $150~mW$, (c) $300~mW$, (d) $500~mW$ respectively. (e)-(h) experimentally measured probe beam profiles at the far field  ($z = \frac{T}{2}+30~cm$) for $w_{x} = ~150~\mu m$ and $w_{y} = 1200~\mu m$ at average pump power of (e) $30~mW$, (f) $150~mW$, (g) $300~mW$, (h) $500~mW$ respectively.}
    \label{fig:4}
\end{figure}

In order to carry out the experimental investigation, we used CL of focal length $f=500~mm$ which resulted in pump beam waist of $w_{x}\approx500~\mu m$ and $w_{y}\approx 1200~\mu m$. Figure 4(a)-(d) shows the recorded probe beam structure at different pump power ($P_p$) levels. It is apparent that the transmitted probe beam undergoes a nominal asymmetric distortion when average incident pump power changes from $P_{p}=30~mW$ to $P_{p}=500~mW$. This could be qualitatively compared with the simulated probe beam profiles in Fig. 2(a)-(d) where probe beam distortion is apparent. Subsequently, we employed a CL of focal length $f=100~mm$ which resulted in $w_{x} \approx 150~\mu m$ and $w_{y} \approx 1200~\mu m$ at focus (ellipticity $\approx~0.12$) and the recorded probe beams are shown in Fig. 4(e)-(h). At an average pump power $P_{p}\geq 300~mW$, there is a discernible splitting of the probe beam into two crescent-shaped lobes and the separation between the lobes increase as a function of pump power $P_p$. When $P_p \approx 500~mW$, the separation between lobes were recorded to be $\approx 1.6~mm$ on the CCD camera plane. The distance between exit face of the cuvette and the CCD plane is $\approx30~cm$. The diffraction limited tracing of the probe beam allows us to ascertain the separation of lobes at the cuvette exit plane which turns out to be $\approx0.15~mm$. This closely resembles the separation obtained from the simulated probe beam profile shown in Fig. 2(h). 

In order to investigate the topological features associated with spatial beam structuring, we follow a two step procedure, namely (i) setting up a balanced Mach-Zehnder interferometer (MZI) which is followed by (ii) an appropriate phase retrieval algorithm for obtaining the acquired geometric phase \cite{Goyal:24,Singh:16}. The MZI is slightly misaligned at the combining non-polarizing beam splitter for obtaining straight line fringes. The reference and the signal probe intensities are subtracted from the recorded fringe pattern at different pump powers 
that are shown in Figs. 5(a)-(d). When $P_p \leq 150~mW$ (with ellipticity $\approx 0.12$), the phase distribution across the wavefront remains uniform. The modification in fringe pattern is discernible at high pump power and the lobes are distinguishable at $P_p \approx 500 ~mW$. A Fourier Transform-based phase retrieval algorithm is employed for extracting phase information from fringe pattern \cite{Takeda:82,Singh:16}. In this method, we select one of the non-zero localized spatial frequency components for recreating the fringe pattern and the incumbent process of computing the inverse Fourier transform allows the recovery of phase distribution across the wavefront. Additionally, we employ transport of intensity (TIE) based algorithm for unwrapping the retrieved phase from the previous phase (see SI document for more details) \cite{Pandey:16,10.1093/jmicro/dfi024}. Figures 5(e)-(h) shows the evolution of unwrapped phase at different pump powers where the black dotted circle defines the computational domain. It is apparent that there is a phase redistribution at $P_p \approx 500~mW$ and the probe beam could be distinctly divided into two regions whose phase bear opposite sign. With respect to Fig. 4(h), it could be observed that the two crescent shaped lobes have opposite phases which epitomizes THE in an all-optical NLO system.  
\begin{figure}
    \centering
    \includegraphics[width=\linewidth]{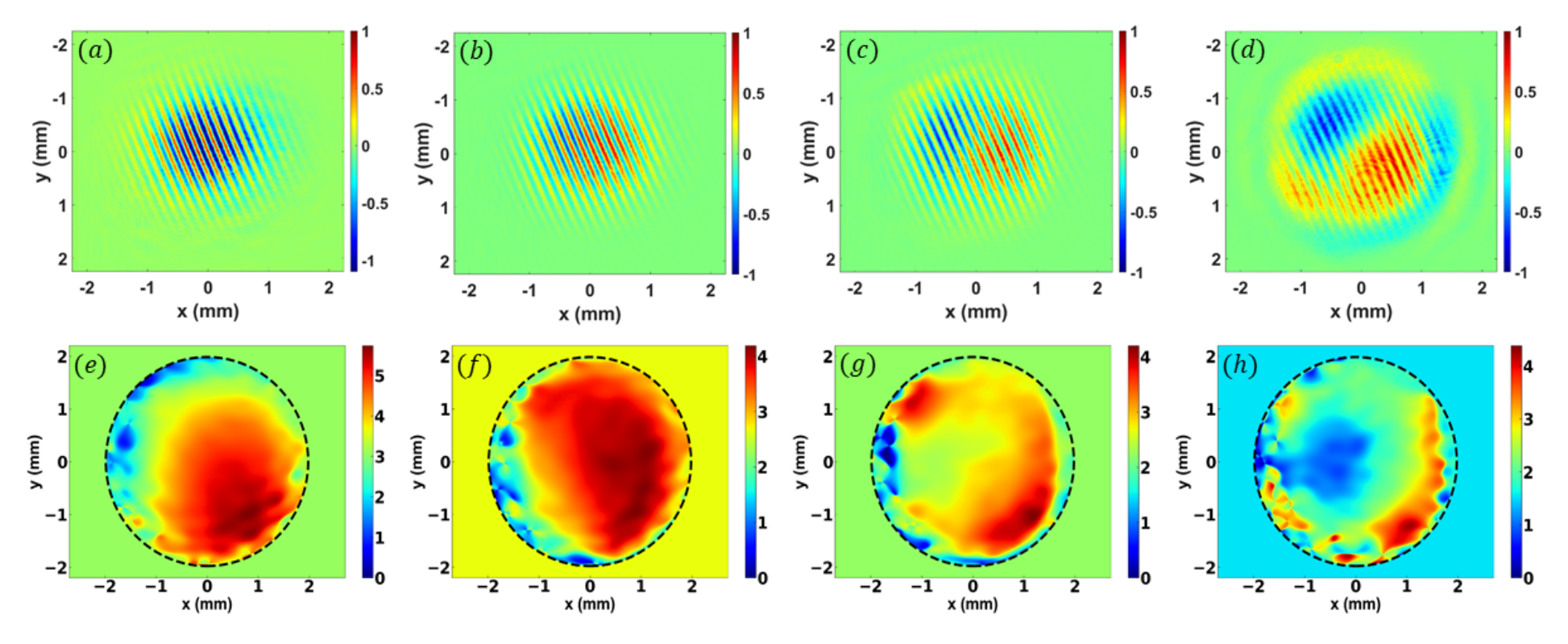}
    \caption{(a)-(d) represents the subtracted pattern of interfering beams from the interference pattern (see supplementary) for $w_{x} = ~150~\mu m$ and $w_{y} = 1200~\mu m$ at average pump power of (a) $30~mW$, (b) $150~mW$, (c) $300~mW$, (d) $500~mW$ respectively. (e)-(h) represents computationally retrieved unwrapped phase profile from the subtracted fringe pattern, for $w_{x} = ~150~\mu m$ and $w_{y} = 1200~\mu m$ at average pump power of (e) $30~mW$, (f) $150~mW$, (g) $300~mW$, (h) $500~mW$ respectively.}
    \label{fig:5}
\end{figure}

In conclusion, we presented a framework for understanding the dynamics of a $\chi^{(3)}$ nonlinear optical interaction from a spin-transport perspective and consequently, we drew a connection between probe beam restructuring with geometric manifestations of the dynamical interaction. In order to realize non-trivial topological manifestation, a non-zero synthetic magnetic field is constructed from an elliptic Gaussian pump beam which strengthened for high pump power. This is accompanied by a $z$-dominant synthetic magnetization which leads to topologically non-trivial splitting of probe beam. The crescent shaped lobes are equivalent to topologically distinguishable bands in space that appear due to THE. The experimental investigation and the phase retrieval of the recorded spatial bands elucidates the formation of topological bands. A full quantum mechanical treatment of the $\chi^{(3)}$ based interaction will provide a feasible scheme for utilizing the spatial beam restructuring for quantum information transport and processing. Since, $\chi^{(3)}$ is relatively small for liquids and solids, a near-resonant self-action effect in a thermal vapor, or preferably a cold-atom ensemble, would be a more viable system for observing such effects at single-photon level.

\section*{Supplementary Material}
See supplementary document-I

\begin{acknowledgments}
S.N. and A.G. are grateful for their fellowships from the Department of Atomic Energy and the Department of Science and Technology, India respectively.
\end{acknowledgments}

\section*{AUTHOR DECLARATIONS}
\subsection*{Conflict of Interest}
The authors have no conflicts to disclose.
\subsection*{Author Contributions}
\textbf{Soumik Nandi:} Conceptualization (equal); Data curation (lead); Formal analysis (equal); Investigation (lead); Methodology (lead); Validation (lead); Visualization (equal); Writing – original draft (equal). \textbf{Arannya Ghosh:} Formal analysis (equal); Resources (equal); Validation (supporting); Visualization (supporting); Writing – original draft (equal). \textbf{Ashok K Mohapatra:} Resources (supporting); Visualization (supporting); Writing – review \& editing (supporting). \textbf{Ritwick Das:} Conceptualization (lead); Funding acquisition (lead); Investigation (supporting); Methodology (equal); Project administration (lead); Resources (lead); Supervision (lead); Validation (supporting); Visualization (equal); Writing – review \& editing (lead).
\section*{Data Availability}
The data that support the findings of this study are available from the corresponding author upon reasonable request.

\section*{References}
\bibliography{aipsamp}

\begin{thebibliography}{28}%
\makeatletter
\providecommand \@ifxundefined [1]{%
 \@ifx{#1\undefined}
}%
\providecommand \@ifnum [1]{%
 \ifnum #1\expandafter \@firstoftwo
 \else \expandafter \@secondoftwo
 \fi
}%
\providecommand \@ifx [1]{%
 \ifx #1\expandafter \@firstoftwo
 \else \expandafter \@secondoftwo
 \fi
}%
\providecommand \natexlab [1]{#1}%
\providecommand \enquote  [1]{``#1''}%
\providecommand \bibnamefont  [1]{#1}%
\providecommand \bibfnamefont [1]{#1}%
\providecommand \citenamefont [1]{#1}%
\providecommand \href@noop [0]{\@secondoftwo}%
\providecommand \href [0]{\begingroup \@sanitize@url \@href}%
\providecommand \@href[1]{\@@startlink{#1}\@@href}%
\providecommand \@@href[1]{\endgroup#1\@@endlink}%
\providecommand \@sanitize@url [0]{\catcode `\\12\catcode `\$12\catcode `\&12\catcode `\#12\catcode `\^12\catcode `\_12\catcode `\%12\relax}%
\providecommand \@@startlink[1]{}%
\providecommand \@@endlink[0]{}%
\providecommand \url  [0]{\begingroup\@sanitize@url \@url }%
\providecommand \@url [1]{\endgroup\@href {#1}{\urlprefix }}%
\providecommand \urlprefix  [0]{URL }%
\providecommand \Eprint [0]{\href }%
\providecommand \doibase [0]{http://dx.doi.org/}%
\providecommand \selectlanguage [0]{\@gobble}%
\providecommand \bibinfo  [0]{\@secondoftwo}%
\providecommand \bibfield  [0]{\@secondoftwo}%
\providecommand \translation [1]{[#1]}%
\providecommand \BibitemOpen [0]{}%
\providecommand \bibitemStop [0]{}%
\providecommand \bibitemNoStop [0]{.\EOS\space}%
\providecommand \EOS [0]{\spacefactor3000\relax}%
\providecommand \BibitemShut  [1]{\csname bibitem#1\endcsname}%
\let\auto@bib@innerbib\@empty
\bibitem [{\citenamefont {Oeckl}(2001)}]{OECKL2001233}%
  \BibitemOpen
  \bibfield  {author} {\bibinfo {author} {\bibfnamefont {R.}~\bibnamefont {Oeckl}},\ }\bibfield  {title} {\enquote {\bibinfo {title} {The quantum geometry of spin and statistics},}\ }\href {\doibase https://doi.org/10.1016/S0393-0440(01)00016-X} {\bibfield  {journal} {\bibinfo  {journal} {Journal of Geometry and Physics}\ }\textbf {\bibinfo {volume} {39}},\ \bibinfo {pages} {233--252} (\bibinfo {year} {2001})}\BibitemShut {NoStop}%
\bibitem [{\citenamefont {Dirac}\ and\ \citenamefont {Polkinghorne}(1958)}]{10.1063/1.3062610}%
  \BibitemOpen
  \bibfield  {author} {\bibinfo {author} {\bibfnamefont {P.~A.~M.}\ \bibnamefont {Dirac}}\ and\ \bibinfo {author} {\bibfnamefont {J.~C.}\ \bibnamefont {Polkinghorne}},\ }\bibfield  {title} {\enquote {\bibinfo {title} {The principles of quantum mechanics},}\ }\href {\doibase 10.1063/1.3062610} {\bibfield  {journal} {\bibinfo  {journal} {Physics Today}\ }\textbf {\bibinfo {volume} {11}},\ \bibinfo {pages} {32--33} (\bibinfo {year} {1958})}\BibitemShut {NoStop}%
\bibitem [{\citenamefont {Gerlach}\ and\ \citenamefont {Stern}(1922)}]{gerlach_experimentelle_1922}%
  \BibitemOpen
  \bibfield  {author} {\bibinfo {author} {\bibfnamefont {W.}~\bibnamefont {Gerlach}}\ and\ \bibinfo {author} {\bibfnamefont {O.}~\bibnamefont {Stern}},\ }\bibfield  {title} {\enquote {\bibinfo {title} {Der experimentelle {Nachweis} der {Richtungsquantelung} im {Magnetfeld}},}\ }\href {\doibase 10.1007/BF01326983} {\bibfield  {journal} {\bibinfo  {journal} {Zeitschrift fur Physik}\ }\textbf {\bibinfo {volume} {9}},\ \bibinfo {pages} {349--352} (\bibinfo {year} {1922})}\BibitemShut {NoStop}%
\bibitem [{\citenamefont {Karpa}\ and\ \citenamefont {Weitz}(2006)}]{Karpa2006}%
  \BibitemOpen
  \bibfield  {author} {\bibinfo {author} {\bibfnamefont {L.}~\bibnamefont {Karpa}}\ and\ \bibinfo {author} {\bibfnamefont {M.}~\bibnamefont {Weitz}},\ }\bibfield  {title} {\enquote {\bibinfo {title} {A stern--gerlach experiment for slow light},}\ }\href {\doibase 10.1038/nphys284} {\bibfield  {journal} {\bibinfo  {journal} {Nature Physics}\ }\textbf {\bibinfo {volume} {2}},\ \bibinfo {pages} {332--335} (\bibinfo {year} {2006})}\BibitemShut {NoStop}%
\bibitem [{\citenamefont {Carusotto}\ and\ \citenamefont {Ciuti}(2013)}]{RevModPhys.85.299}%
  \BibitemOpen
  \bibfield  {author} {\bibinfo {author} {\bibfnamefont {I.}~\bibnamefont {Carusotto}}\ and\ \bibinfo {author} {\bibfnamefont {C.}~\bibnamefont {Ciuti}},\ }\bibfield  {title} {\enquote {\bibinfo {title} {Quantum fluids of light},}\ }\href {\doibase 10.1103/RevModPhys.85.299} {\bibfield  {journal} {\bibinfo  {journal} {Rev. Mod. Phys.}\ }\textbf {\bibinfo {volume} {85}},\ \bibinfo {pages} {299--366} (\bibinfo {year} {2013})}\BibitemShut {NoStop}%
\bibitem [{\citenamefont {\ifmmode \check{Z}\else \v{Z}\fi{}uti\ifmmode~\acute{c}\else \'{c}\fi{}}, \citenamefont {Fabian},\ and\ \citenamefont {Das~Sarma}(2004)}]{RevModPhys.76.323}%
  \BibitemOpen
  \bibfield  {author} {\bibinfo {author} {\bibfnamefont {I.}~\bibnamefont {\ifmmode \check{Z}\else \v{Z}\fi{}uti\ifmmode~\acute{c}\else \'{c}\fi{}}}, \bibinfo {author} {\bibfnamefont {J.}~\bibnamefont {Fabian}}, \ and\ \bibinfo {author} {\bibfnamefont {S.}~\bibnamefont {Das~Sarma}},\ }\bibfield  {title} {\enquote {\bibinfo {title} {Spintronics: Fundamentals and applications},}\ }\href {\doibase 10.1103/RevModPhys.76.323} {\bibfield  {journal} {\bibinfo  {journal} {Rev. Mod. Phys.}\ }\textbf {\bibinfo {volume} {76}},\ \bibinfo {pages} {323--410} (\bibinfo {year} {2004})}\BibitemShut {NoStop}%
\bibitem [{\citenamefont {Parkin}\ \emph {et~al.}(1999)\citenamefont {Parkin}, \citenamefont {Roche}, \citenamefont {Samant}, \citenamefont {Rice}, \citenamefont {Beyers}, \citenamefont {Scheuerlein}, \citenamefont {O’Sullivan}, \citenamefont {Brown}, \citenamefont {Bucchigano}, \citenamefont {Abraham}, \citenamefont {Lu}, \citenamefont {Rooks}, \citenamefont {Trouilloud}, \citenamefont {Wanner},\ and\ \citenamefont {Gallagher}}]{10.1063/1.369932}%
  \BibitemOpen
  \bibfield  {author} {\bibinfo {author} {\bibfnamefont {S.~S.~P.}\ \bibnamefont {Parkin}}, \bibinfo {author} {\bibfnamefont {K.~P.}\ \bibnamefont {Roche}}, \bibinfo {author} {\bibfnamefont {M.~G.}\ \bibnamefont {Samant}}, \bibinfo {author} {\bibfnamefont {P.~M.}\ \bibnamefont {Rice}}, \bibinfo {author} {\bibfnamefont {R.~B.}\ \bibnamefont {Beyers}}, \bibinfo {author} {\bibfnamefont {R.~E.}\ \bibnamefont {Scheuerlein}}, \bibinfo {author} {\bibfnamefont {E.~J.}\ \bibnamefont {O’Sullivan}}, \bibinfo {author} {\bibfnamefont {S.~L.}\ \bibnamefont {Brown}}, \bibinfo {author} {\bibfnamefont {J.}~\bibnamefont {Bucchigano}}, \bibinfo {author} {\bibfnamefont {D.~W.}\ \bibnamefont {Abraham}}, \bibinfo {author} {\bibfnamefont {Y.}~\bibnamefont {Lu}}, \bibinfo {author} {\bibfnamefont {M.}~\bibnamefont {Rooks}}, \bibinfo {author} {\bibfnamefont {P.~L.}\ \bibnamefont {Trouilloud}}, \bibinfo {author} {\bibfnamefont {R.~A.}\ \bibnamefont {Wanner}}, \ and\ \bibinfo {author} {\bibfnamefont {W.~J.}\ \bibnamefont
  {Gallagher}},\ }\bibfield  {title} {\enquote {\bibinfo {title} {Exchange-biased magnetic tunnel junctions and application to nonvolatile magnetic random access memory (invited)},}\ }\href {\doibase 10.1063/1.369932} {\bibfield  {journal} {\bibinfo  {journal} {Journal of Applied Physics}\ }\textbf {\bibinfo {volume} {85}},\ \bibinfo {pages} {5828--5833} (\bibinfo {year} {1999})}\BibitemShut {NoStop}%
\bibitem [{\citenamefont {Leitao}\ \emph {et~al.}(2024)\citenamefont {Leitao}, \citenamefont {Riel}, \citenamefont {Rasly}, \citenamefont {Araujo}, \citenamefont {Salvador}, \citenamefont {Paz},\ and\ \citenamefont {Koopmans}}]{leitao_enhanced_2024}%
  \BibitemOpen
  \bibfield  {author} {\bibinfo {author} {\bibfnamefont {D.~C.}\ \bibnamefont {Leitao}}, \bibinfo {author} {\bibfnamefont {F.~J. F.~V.}\ \bibnamefont {Riel}}, \bibinfo {author} {\bibfnamefont {M.}~\bibnamefont {Rasly}}, \bibinfo {author} {\bibfnamefont {P.~D.~R.}\ \bibnamefont {Araujo}}, \bibinfo {author} {\bibfnamefont {M.}~\bibnamefont {Salvador}}, \bibinfo {author} {\bibfnamefont {E.}~\bibnamefont {Paz}}, \ and\ \bibinfo {author} {\bibfnamefont {B.}~\bibnamefont {Koopmans}},\ }\bibfield  {title} {\enquote {\bibinfo {title} {Enhanced performance and functionality in spintronic sensors},}\ }\href {\doibase 10.1038/s44306-024-00058-9} {\bibfield  {journal} {\bibinfo  {journal} {npj Spintronics}\ }\textbf {\bibinfo {volume} {2}},\ \bibinfo {pages} {54} (\bibinfo {year} {2024})}\BibitemShut {NoStop}%
\bibitem [{\citenamefont {Li}\ \emph {et~al.}(2019)\citenamefont {Li}, \citenamefont {Shi}, \citenamefont {Lei}, \citenamefont {He},\ and\ \citenamefont {Liu}}]{8951547}%
  \BibitemOpen
  \bibfield  {author} {\bibinfo {author} {\bibfnamefont {T.}~\bibnamefont {Li}}, \bibinfo {author} {\bibfnamefont {Q.}~\bibnamefont {Shi}}, \bibinfo {author} {\bibfnamefont {Z.}~\bibnamefont {Lei}}, \bibinfo {author} {\bibfnamefont {L.}~\bibnamefont {He}}, \ and\ \bibinfo {author} {\bibfnamefont {B.}~\bibnamefont {Liu}},\ }\bibfield  {title} {\enquote {\bibinfo {title} {Research on mechanism and key technology of intelligent vehicles brake by wire system},}\ }in\ \href {\doibase 10.1109/CVCI47823.2019.8951547} {\emph {\bibinfo {booktitle} {2019 3rd Conference on Vehicle Control and Intelligence (CVCI)}}}\ (\bibinfo {year} {2019})\ pp.\ \bibinfo {pages} {1--8}\BibitemShut {NoStop}%
\bibitem [{\citenamefont {Parkin}, \citenamefont {Hayashi},\ and\ \citenamefont {Thomas}(2008)}]{doi:10.1126/science.1145799}%
  \BibitemOpen
  \bibfield  {author} {\bibinfo {author} {\bibfnamefont {S.~S.~P.}\ \bibnamefont {Parkin}}, \bibinfo {author} {\bibfnamefont {M.}~\bibnamefont {Hayashi}}, \ and\ \bibinfo {author} {\bibfnamefont {L.}~\bibnamefont {Thomas}},\ }\bibfield  {title} {\enquote {\bibinfo {title} {Magnetic domain-wall racetrack memory},}\ }\href {\doibase 10.1126/science.1145799} {\bibfield  {journal} {\bibinfo  {journal} {Science}\ }\textbf {\bibinfo {volume} {320}},\ \bibinfo {pages} {190--194} (\bibinfo {year} {2008})}\BibitemShut {NoStop}%
\bibitem [{\citenamefont {Ghosh}(2016)}]{7559746}%
  \BibitemOpen
  \bibfield  {author} {\bibinfo {author} {\bibfnamefont {S.}~\bibnamefont {Ghosh}},\ }\bibfield  {title} {\enquote {\bibinfo {title} {Spintronics and security: Prospects, vulnerabilities, attack models, and preventions},}\ }\href {\doibase 10.1109/JPROC.2016.2583419} {\bibfield  {journal} {\bibinfo  {journal} {Proceedings of the IEEE}\ }\textbf {\bibinfo {volume} {104}},\ \bibinfo {pages} {1864--1893} (\bibinfo {year} {2016})}\BibitemShut {NoStop}%
\bibitem [{\citenamefont {Yang}\ \emph {et~al.}(2024)\citenamefont {Yang}, \citenamefont {Zhao}, \citenamefont {Zhang}, \citenamefont {Xing}, \citenamefont {Du}, \citenamefont {Li}, \citenamefont {Mochizuki}, \citenamefont {Xu}, \citenamefont {Åkerman},\ and\ \citenamefont {Zhou}}]{10.1063/5.0218280}%
  \BibitemOpen
  \bibfield  {author} {\bibinfo {author} {\bibfnamefont {S.}~\bibnamefont {Yang}}, \bibinfo {author} {\bibfnamefont {Y.}~\bibnamefont {Zhao}}, \bibinfo {author} {\bibfnamefont {X.}~\bibnamefont {Zhang}}, \bibinfo {author} {\bibfnamefont {X.}~\bibnamefont {Xing}}, \bibinfo {author} {\bibfnamefont {H.}~\bibnamefont {Du}}, \bibinfo {author} {\bibfnamefont {X.}~\bibnamefont {Li}}, \bibinfo {author} {\bibfnamefont {M.}~\bibnamefont {Mochizuki}}, \bibinfo {author} {\bibfnamefont {X.}~\bibnamefont {Xu}}, \bibinfo {author} {\bibfnamefont {J.}~\bibnamefont {Åkerman}}, \ and\ \bibinfo {author} {\bibfnamefont {Y.}~\bibnamefont {Zhou}},\ }\bibfield  {title} {\enquote {\bibinfo {title} {Fundamentals and applications of the skyrmion hall effect},}\ }\href {\doibase 10.1063/5.0218280} {\bibfield  {journal} {\bibinfo  {journal} {Applied Physics Reviews}\ }\textbf {\bibinfo {volume} {11}},\ \bibinfo {pages} {041335} (\bibinfo {year} {2024})}\BibitemShut {NoStop}%
\bibitem [{\citenamefont {Bashan}\ \emph {et~al.}(2025)\citenamefont {Bashan}, \citenamefont {Izhak}, \citenamefont {Yesharim},\ and\ \citenamefont {Arie}}]{10.1063/5.0241546}%
  \BibitemOpen
  \bibfield  {author} {\bibinfo {author} {\bibfnamefont {G.}~\bibnamefont {Bashan}}, \bibinfo {author} {\bibfnamefont {S.}~\bibnamefont {Izhak}}, \bibinfo {author} {\bibfnamefont {O.}~\bibnamefont {Yesharim}}, \ and\ \bibinfo {author} {\bibfnamefont {A.}~\bibnamefont {Arie}},\ }\bibfield  {title} {\enquote {\bibinfo {title} {Spintronic effects and devices in nonlinear optics},}\ }\href {\doibase 10.1063/5.0241546} {\bibfield  {journal} {\bibinfo  {journal} {APL Photonics}\ }\textbf {\bibinfo {volume} {10}},\ \bibinfo {pages} {010904} (\bibinfo {year} {2025})}\BibitemShut {NoStop}%
\bibitem [{\citenamefont {Liu}\ \emph {et~al.}(2015)\citenamefont {Liu}, \citenamefont {Wang}, \citenamefont {Xiao}, \citenamefont {Hang},\ and\ \citenamefont {Li}}]{10.1063/1.4938003}%
  \BibitemOpen
  \bibfield  {author} {\bibinfo {author} {\bibfnamefont {F.}~\bibnamefont {Liu}}, \bibinfo {author} {\bibfnamefont {S.}~\bibnamefont {Wang}}, \bibinfo {author} {\bibfnamefont {S.}~\bibnamefont {Xiao}}, \bibinfo {author} {\bibfnamefont {Z.~H.}\ \bibnamefont {Hang}}, \ and\ \bibinfo {author} {\bibfnamefont {J.}~\bibnamefont {Li}},\ }\bibfield  {title} {\enquote {\bibinfo {title} {Polarization-dependent optics using gauge-field metamaterials},}\ }\href {\doibase 10.1063/1.4938003} {\bibfield  {journal} {\bibinfo  {journal} {Applied Physics Letters}\ }\textbf {\bibinfo {volume} {107}},\ \bibinfo {pages} {241106} (\bibinfo {year} {2015})}\BibitemShut {NoStop}%
\bibitem [{\citenamefont {Karnieli}\ \emph {et~al.}(2021)\citenamefont {Karnieli}, \citenamefont {Tsesses}, \citenamefont {Bartal},\ and\ \citenamefont {Arie}}]{Karnieli2021}%
  \BibitemOpen
  \bibfield  {author} {\bibinfo {author} {\bibfnamefont {A.}~\bibnamefont {Karnieli}}, \bibinfo {author} {\bibfnamefont {S.}~\bibnamefont {Tsesses}}, \bibinfo {author} {\bibfnamefont {G.}~\bibnamefont {Bartal}}, \ and\ \bibinfo {author} {\bibfnamefont {A.}~\bibnamefont {Arie}},\ }\bibfield  {title} {\enquote {\bibinfo {title} {Emulating spin transport with nonlinear optics, from high-order skyrmions to the topological hall effect},}\ }\href {\doibase 10.1038/s41467-021-21250-z} {\bibfield  {journal} {\bibinfo  {journal} {Nature Communications}\ }\textbf {\bibinfo {volume} {12}},\ \bibinfo {pages} {1092} (\bibinfo {year} {2021})}\BibitemShut {NoStop}%
\bibitem [{\citenamefont {Izhak}\ \emph {et~al.}(2024)\citenamefont {Izhak}, \citenamefont {Karnieli}, \citenamefont {Yesharim}, \citenamefont {Tsesses},\ and\ \citenamefont {Arie}}]{Izhak:24}%
  \BibitemOpen
  \bibfield  {author} {\bibinfo {author} {\bibfnamefont {S.}~\bibnamefont {Izhak}}, \bibinfo {author} {\bibfnamefont {A.}~\bibnamefont {Karnieli}}, \bibinfo {author} {\bibfnamefont {O.}~\bibnamefont {Yesharim}}, \bibinfo {author} {\bibfnamefont {S.}~\bibnamefont {Tsesses}}, \ and\ \bibinfo {author} {\bibfnamefont {A.}~\bibnamefont {Arie}},\ }\bibfield  {title} {\enquote {\bibinfo {title} {All-optical spin valve effect in nonlinear optics},}\ }\href {\doibase 10.1364/OL.517794} {\bibfield  {journal} {\bibinfo  {journal} {Opt. Lett.}\ }\textbf {\bibinfo {volume} {49}},\ \bibinfo {pages} {1025--1028} (\bibinfo {year} {2024})}\BibitemShut {NoStop}%
\bibitem [{\citenamefont {Jain}\ \emph {et~al.}(2025)\citenamefont {Jain}, \citenamefont {Hlaing}, \citenamefont {Fan}, \citenamefont {Midkiff}, \citenamefont {Ning}, \citenamefont {Feng}, \citenamefont {Hsiao}, \citenamefont {Camp},\ and\ \citenamefont {Chen}}]{10.1063/5.0218564}%
  \BibitemOpen
  \bibfield  {author} {\bibinfo {author} {\bibfnamefont {S.}~\bibnamefont {Jain}}, \bibinfo {author} {\bibfnamefont {M.~H.}\ \bibnamefont {Hlaing}}, \bibinfo {author} {\bibfnamefont {K.-C.}\ \bibnamefont {Fan}}, \bibinfo {author} {\bibfnamefont {J.}~\bibnamefont {Midkiff}}, \bibinfo {author} {\bibfnamefont {S.}~\bibnamefont {Ning}}, \bibinfo {author} {\bibfnamefont {C.}~\bibnamefont {Feng}}, \bibinfo {author} {\bibfnamefont {P.-Y.}\ \bibnamefont {Hsiao}}, \bibinfo {author} {\bibfnamefont {P.~T.}\ \bibnamefont {Camp}}, \ and\ \bibinfo {author} {\bibfnamefont {R.~T.}\ \bibnamefont {Chen}},\ }\bibfield  {title} {\enquote {\bibinfo {title} {Incubating advances in integrated photonics with emerging sensing and computational capabilities},}\ }\href {\doibase 10.1063/5.0218564} {\bibfield  {journal} {\bibinfo  {journal} {Applied Physics Reviews}\ }\textbf {\bibinfo {volume} {12}},\ \bibinfo {pages} {011337} (\bibinfo {year} {2025})}\BibitemShut {NoStop}%
\bibitem [{\citenamefont {Westerberg}\ \emph {et~al.}(2016)\citenamefont {Westerberg}, \citenamefont {Maitland}, \citenamefont {Faccio}, \citenamefont {Wilson}, \citenamefont {\"Ohberg},\ and\ \citenamefont {Wright}}]{PhysRevA.94.023805}%
  \BibitemOpen
  \bibfield  {author} {\bibinfo {author} {\bibfnamefont {N.}~\bibnamefont {Westerberg}}, \bibinfo {author} {\bibfnamefont {C.}~\bibnamefont {Maitland}}, \bibinfo {author} {\bibfnamefont {D.}~\bibnamefont {Faccio}}, \bibinfo {author} {\bibfnamefont {K.}~\bibnamefont {Wilson}}, \bibinfo {author} {\bibfnamefont {P.}~\bibnamefont {\"Ohberg}}, \ and\ \bibinfo {author} {\bibfnamefont {E.~M.}\ \bibnamefont {Wright}},\ }\bibfield  {title} {\enquote {\bibinfo {title} {Synthetic magnetism for photon fluids},}\ }\href {\doibase 10.1103/PhysRevA.94.023805} {\bibfield  {journal} {\bibinfo  {journal} {Phys. Rev. A}\ }\textbf {\bibinfo {volume} {94}},\ \bibinfo {pages} {023805} (\bibinfo {year} {2016})}\BibitemShut {NoStop}%
\bibitem [{\citenamefont {Bender}\ and\ \citenamefont {Boettcher}(1998)}]{PhysRevLett.80.5243}%
  \BibitemOpen
  \bibfield  {author} {\bibinfo {author} {\bibfnamefont {C.~M.}\ \bibnamefont {Bender}}\ and\ \bibinfo {author} {\bibfnamefont {S.}~\bibnamefont {Boettcher}},\ }\bibfield  {title} {\enquote {\bibinfo {title} {Real spectra in non-hermitian hamiltonians having ${P}{T}$ symmetry},}\ }\href {\doibase 10.1103/PhysRevLett.80.5243} {\bibfield  {journal} {\bibinfo  {journal} {Phys. Rev. Lett.}\ }\textbf {\bibinfo {volume} {80}},\ \bibinfo {pages} {5243--5246} (\bibinfo {year} {1998})}\BibitemShut {NoStop}%
\bibitem [{\citenamefont {Bender}, \citenamefont {Brody},\ and\ \citenamefont {Jones}(2002)}]{PhysRevLett.89.270401}%
  \BibitemOpen
  \bibfield  {author} {\bibinfo {author} {\bibfnamefont {C.~M.}\ \bibnamefont {Bender}}, \bibinfo {author} {\bibfnamefont {D.~C.}\ \bibnamefont {Brody}}, \ and\ \bibinfo {author} {\bibfnamefont {H.~F.}\ \bibnamefont {Jones}},\ }\bibfield  {title} {\enquote {\bibinfo {title} {Complex extension of quantum mechanics},}\ }\href {\doibase 10.1103/PhysRevLett.89.270401} {\bibfield  {journal} {\bibinfo  {journal} {Phys. Rev. Lett.}\ }\textbf {\bibinfo {volume} {89}},\ \bibinfo {pages} {270401} (\bibinfo {year} {2002})}\BibitemShut {NoStop}%
\bibitem [{\citenamefont {Boyd}(2008)}]{boyd_nonlinear_2008}%
  \BibitemOpen
  \bibfield  {author} {\bibinfo {author} {\bibfnamefont {R.~W.}\ \bibnamefont {Boyd}},\ }\href@noop {} {\emph {\bibinfo {title} {Nonlinear {Optics}}}},\ \bibinfo {edition} {3rd}\ ed.\ (\bibinfo  {publisher} {Academic Press, Inc.},\ \bibinfo {address} {USA},\ \bibinfo {year} {2008})\BibitemShut {NoStop}%
\bibitem [{\citenamefont {Nandi}\ \emph {et~al.}(2025)\citenamefont {Nandi}, \citenamefont {Ghosh}, \citenamefont {Beniwal}, \citenamefont {Mondal}, \citenamefont {Mohapatra},\ and\ \citenamefont {Das}}]{Nandi:25}%
  \BibitemOpen
  \bibfield  {author} {\bibinfo {author} {\bibfnamefont {S.}~\bibnamefont {Nandi}}, \bibinfo {author} {\bibfnamefont {A.}~\bibnamefont {Ghosh}}, \bibinfo {author} {\bibfnamefont {D.}~\bibnamefont {Beniwal}}, \bibinfo {author} {\bibfnamefont {A.}~\bibnamefont {Mondal}}, \bibinfo {author} {\bibfnamefont {A.~K.}\ \bibnamefont {Mohapatra}}, \ and\ \bibinfo {author} {\bibfnamefont {R.}~\bibnamefont {Das}},\ }\bibfield  {title} {\enquote {\bibinfo {title} {All-optical stern-gerlach effect in a parity-time anti-symmetric nonlinear refractive medium},}\ }\href {\doibase 10.1364/OE.557624} {\bibfield  {journal} {\bibinfo  {journal} {Opt. Express}\ }\textbf {\bibinfo {volume} {33}},\ \bibinfo {pages} {24237--24248} (\bibinfo {year} {2025})}\BibitemShut {NoStop}%
\bibitem [{\citenamefont {Everschor-Sitte}\ and\ \citenamefont {Sitte}(2014)}]{10.1063/1.4870695}%
  \BibitemOpen
  \bibfield  {author} {\bibinfo {author} {\bibfnamefont {K.}~\bibnamefont {Everschor-Sitte}}\ and\ \bibinfo {author} {\bibfnamefont {M.}~\bibnamefont {Sitte}},\ }\bibfield  {title} {\enquote {\bibinfo {title} {Real-space berry phases: Skyrmion soccer (invited)},}\ }\href {\doibase 10.1063/1.4870695} {\bibfield  {journal} {\bibinfo  {journal} {Journal of Applied Physics}\ }\textbf {\bibinfo {volume} {115}},\ \bibinfo {pages} {172602} (\bibinfo {year} {2014})}\BibitemShut {NoStop}%
\bibitem [{\citenamefont {Goyal}\ and\ \citenamefont {Khare}(2024)}]{Goyal:24}%
  \BibitemOpen
  \bibfield  {author} {\bibinfo {author} {\bibfnamefont {N.}~\bibnamefont {Goyal}}\ and\ \bibinfo {author} {\bibfnamefont {K.}~\bibnamefont {Khare}},\ }\bibfield  {title} {\enquote {\bibinfo {title} {Carrier-frequency estimation for digital holograms of phase objects},}\ }\href {\doibase 10.1364/AO.505663} {\bibfield  {journal} {\bibinfo  {journal} {Appl. Opt.}\ }\textbf {\bibinfo {volume} {63}},\ \bibinfo {pages} {B42--B48} (\bibinfo {year} {2024})}\BibitemShut {NoStop}%
\bibitem [{\citenamefont {Singh}\ and\ \citenamefont {Khare}(2016)}]{Singh:16}%
  \BibitemOpen
  \bibfield  {author} {\bibinfo {author} {\bibfnamefont {M.}~\bibnamefont {Singh}}\ and\ \bibinfo {author} {\bibfnamefont {K.}~\bibnamefont {Khare}},\ }\bibfield  {title} {\enquote {\bibinfo {title} {Accurate efficient carrier estimation for single-shot digital holographic imaging},}\ }\href {\doibase 10.1364/OL.41.004871} {\bibfield  {journal} {\bibinfo  {journal} {Opt. Lett.}\ }\textbf {\bibinfo {volume} {41}},\ \bibinfo {pages} {4871--4874} (\bibinfo {year} {2016})}\BibitemShut {NoStop}%
\bibitem [{\citenamefont {Takeda}, \citenamefont {Ina},\ and\ \citenamefont {Kobayashi}(1982)}]{Takeda:82}%
  \BibitemOpen
  \bibfield  {author} {\bibinfo {author} {\bibfnamefont {M.}~\bibnamefont {Takeda}}, \bibinfo {author} {\bibfnamefont {H.}~\bibnamefont {Ina}}, \ and\ \bibinfo {author} {\bibfnamefont {S.}~\bibnamefont {Kobayashi}},\ }\bibfield  {title} {\enquote {\bibinfo {title} {Fourier-transform method of fringe-pattern analysis for computer-based topography and interferometry},}\ }\href {\doibase 10.1364/JOSA.72.000156} {\bibfield  {journal} {\bibinfo  {journal} {J. Opt. Soc. Am.}\ }\textbf {\bibinfo {volume} {72}},\ \bibinfo {pages} {156--160} (\bibinfo {year} {1982})}\BibitemShut {NoStop}%
\bibitem [{\citenamefont {Pandey}, \citenamefont {Ghosh},\ and\ \citenamefont {Khare}(2016)}]{Pandey:16}%
  \BibitemOpen
  \bibfield  {author} {\bibinfo {author} {\bibfnamefont {N.}~\bibnamefont {Pandey}}, \bibinfo {author} {\bibfnamefont {A.}~\bibnamefont {Ghosh}}, \ and\ \bibinfo {author} {\bibfnamefont {K.}~\bibnamefont {Khare}},\ }\bibfield  {title} {\enquote {\bibinfo {title} {Two-dimensional phase unwrapping using the transport of intensity equation},}\ }\href {\doibase 10.1364/AO.55.002418} {\bibfield  {journal} {\bibinfo  {journal} {Appl. Opt.}\ }\textbf {\bibinfo {volume} {55}},\ \bibinfo {pages} {2418--2425} (\bibinfo {year} {2016})}\BibitemShut {NoStop}%
\bibitem [{\citenamefont {Ishizuka}\ and\ \citenamefont {Allman}(2005)}]{10.1093/jmicro/dfi024}%
  \BibitemOpen
  \bibfield  {author} {\bibinfo {author} {\bibfnamefont {K.}~\bibnamefont {Ishizuka}}\ and\ \bibinfo {author} {\bibfnamefont {B.}~\bibnamefont {Allman}},\ }\bibfield  {title} {\enquote {\bibinfo {title} {Phase measurement of atomic resolution image using transport of intensity equation},}\ }\href {\doibase 10.1093/jmicro/dfi024} {\bibfield  {journal} {\bibinfo  {journal} {Journal of Electron Microscopy}\ }\textbf {\bibinfo {volume} {54}},\ \bibinfo {pages} {191--197} (\bibinfo {year} {2005})}\BibitemShut {NoStop}%
\end{thebibliography}%

\end{document}